\documentclass[preprint,floatfix,aps,prd,showpacs,nofootinbib,amsmath,amssymb,amsfonts,superscriptaddress]{revtex4-1}
\usepackage{graphicx}
\usepackage{color}
\usepackage{dsfont}
\usepackage[normalem]{ulem}
\usepackage{hyperref}
\usepackage{mathtools}
\usepackage{mathrsfs}
\usepackage{calrsfs}
\usepackage{comment}
\usepackage[utf8]{inputenc}
\usepackage{multirow} %extra package used
\usepackage{subfigure}
\usepackage{varioref}
\usepackage[active]{srcltx}

\expandafter\ifx\csname package@font\endcsname\relax\else
\expandafter\expandafter
\expandafter\usepackage
\expandafter\expandafter
\expandafter{\csname package@font\endcsname}%
\fi
\hypersetup{
	colorlinks=true,       % false: boxed links; true: colored links
	linkcolor=blue,          % color of internal links
	citecolor=blue,        % color of links to bibliography
}
\usepackage[section]{placeins}

\usepackage{fancybox}
\labelformat{figure}{Fig.~#1}

\begin{document}
	\title{Helical magnetic fields from Riemann coupling} 
	\author{Ashu Kushwaha} 
	\email{ashu712@iitb.ac.in}
	\affiliation{Department of Physics, Indian Institute of Technology Bombay, Mumbai 400076, India}
	\author{S. Shankaranarayanan}
	\email{shanki@phy.iitb.ac.in}
	\affiliation{Department of Physics, Indian Institute of Technology Bombay, Mumbai 400076, India}
\begin{abstract}
We study the inflationary generation of helical magnetic fields from the Riemann coupling with the electromagnetic field. Most models in the literature introduce non-minimal coupling to the electromagnetic fields with a scalar field,  hence, breaking the conformal invariance.  In this work, we show that non-minimal coupling to the Riemann tensor generates sufficient primordial helical magnetic fields at all observable scales. We explicitly show that one of the helical states decay while the other helical mode increases, leading to a net non-zero helicity. Our model has three key features:~(i) the helical power-spectrum has a slight red-tilt for slow-roll inflation consistent with bounds from observations and free from backreaction problem, (ii) the energy density of the helical fields generated is at least one order of magnitude larger than the scalar-field coupled models, and (iii) unlike the scalar field coupled models, the generated helical fields are insensitive to the reheating dynamics. We show that our model generates the magnetic field of strength $0.01$~PicoGauss over $\rm{Mpc}$ scale.
\end{abstract}
	\maketitle
\section{Introduction}

Effective field theory (EFT) now forms a standard tool in early-Universe cosmology~\cite{2017-Burgess-arXiv,2014-Agarwal.etal-JCAP,2008-Weinberg-PRD,2007-Cheung.etal-JHEP}. Effective field theories rely on the separation of the energy scales of interest for observations and the underlying physics of the early Universe near the singularity~\cite{1994-Donoghue-PRD,1996-Donoghue.etal,2003-Burgess-LivRev}. In an effective field theory approach, the usual requirement of renormalizability is too strong. Instead, it demands that a finite number of parameters describe the physics up to effects suppressed by $(E/M)^n$ where $ E $ is the energy of the particles corresponding to the quantum fields, and $ M $ is the energy scale below which the effective field theory description is valid. As the value of $n$ increases, more parameters are required to describe the Physics using EFT. 

Effective field theory description for gravity has shown that gravity and quantum mechanics can be compatible with the energies that have been experimentally probed~\cite{Donoghue:2015hwa}. EFT has been successfully applied to inflationary cosmology, especially for the single scalar field inflation model were the constants in the higher derivative terms of the effective Lagrangian take values that are powers of $M$, with coefficients roughly of order unity~\cite{2008-Weinberg-PRD,2007-Cheung.etal-JHEP}. However, these EFTs aim to obtain a generic prediction for density perturbations in a single-scalar field inflationary models~\cite{2007-Cheung.etal-JHEP,2019-Bastero-Gil.etal-arXiv}. Hence, the analyses can not be extended to other fields (like electromagnetic fields) during inflation.

Recently, the general effective field theory of gravity coupled to the Standard Model of particle physics was constructed~\cite{2019-Ruhdorfer.etal-JHEP}. The authors systematically showed that the first gravity operators appear at mass dimension $6$ in the series expansion, and these operators only couple to the standard model Bosons. They also showed that (i) no new gravity operators appear at mass dimension $7$, (ii) in mass dimension $8$ the standard model Fermions appear, and (iii) coupling between the scalar (Higgs) field and the standard model gauge Bosons appear \emph{only} at mass dimension $8$. (Note that these corrections do not include higher derivative (Galileon) terms in the electromagnetic action~\cite{2017-Debottam.Shankaranarayanan-JCAP,2019-Kushwaha.Shankaranarayanan-PRD}.)

In this work, we limit to mass dimension $6$ operators coupling to the gauge field, specifically, to the electromagnetic field. We concentrate on the coupling of the electromagnetic field with the dual Riemann tensor $\tilde{R}^{\mu \nu \rho \sigma} ( \equiv \epsilon^{\mu \nu \alpha \beta} R_{\alpha \beta}^{~~~\rho \sigma} / 2$). We show that such a term leads to the generation of a helical magnetic field during inflation~\cite{2001-Vachaspati-PRL}. 

Magnetic fields have been observed at all scales in the Universe; however, there is no compelling model of the origin of large scale magnetic fields. Observations from Faraday rotation and synchrotron radiation show the presence of micro-Gauss strength magnetic fields in the galaxies and the clusters of galaxies~\cite{1994-Kronberg-Rept.Prog.Phys.,2001-Grasso.etal-PhyRep,2001-Grasso.etal-PhyRep,2002-Widrow-Rev.Mod.Phys.,2013-Durrer.Neronov-Arxiv,2016-Subramanian-Arxiv}. 

While the magnetic field measurements from Faraday rotation and synchrotron radiation provide upper bounds of the magnetic fields, the FERMI measurement of gamma-rays emitted by blazars provides a lower bound of the order of $10^{-15}$ G in intergalactic voids~\cite{2010-Neronov.Vovk-Sci}. However, inflation can not generate these large scale magnetic fields as the standard 4-D electromagnetic action is conformally invariant, so the magnetic fields will dilute with the expansion of the universe. To amplify the quantum fluctuations of the electromagnetic fields in the early Universe,  
one needs to break the conformal invariance of the action~\cite{1988-Turner.Widrow-PRD,2002-Widrow-Rev.Mod.Phys.,2013-Durrer.Neronov-Arxiv,2016-Subramanian-Arxiv,2018-Subramanian-arXiv,2014-Tsagas-arXiv}. 
However, it has been argued that in spatially open Friedmann universes these field will survive with significant strength without changing the standard electromagnetism  \cite{2011-Barrow.Tsagas-MNRAS,2008-Barrow.Tsagas-PRD}.

While mechanisms to generate non-helical fields were proposed four decades ago~\cite{1988-Turner.Widrow-PRD,1991-Ratra-Apj.Lett,1993-Dolgov-PRD}, the generation of helical fields is recent~\cite{2001-Vachaspati-PRL,2003-Caprini.etal-PRD,2005-Campanelli-Giannotti-PRD}.
One of the interests in primordial magnetic helicity is that it can be a direct indication of parity violation in the early Universe and may be related to the matter-antimatter asymmetry in the early Universe~\cite{1996-Davidson-PLB}. Besides, the conservation of helicity leads to an inverse cascade in the turbulent plasma era, that can move power from small to large scales~\cite{2003-Caprini.etal-PRD}. Hence, the decay rate of energy density and coherence length is slower than the non-helical fields during these epochs. It has been shown that helical magnetic fields will leave distinct signatures in CMB, such as TE- and EB-cross-correlations~\cite{2006-Kahniashvil-NAR,2018-Chowdhury.etal-JCAP}.

As mentioned above, there has been a lot of interest in generating a primordial helical field in the early Universe~\cite{2001-Vachaspati-PRL,2003-Caprini.etal-PRD,2005-Campanelli-Giannotti-PRD,2018-Sharma.Subramanian.Seshadri.PRD,2009-Caprini.Durrer.Fenu-JCAP,2009-Campanelli-IJMPD,2019-Shtanov-Ukr.PJ}. However, most models introduce non-minimal coupling of the electromagnetic fields with a (pseudo-)scalar. While this leads to the breaking of conformal invariance of the electromagnetic field, due to non-minimal coupling, extra degrees of freedom are present at all energies. More importantly, these extra degrees of freedom propagate even at the low-energy and can potentially lead to the strong coupling problem~\cite{2009-Demozzi.etal-JCAP}. It is possible to overcome the strong-coupling problem for a narrow range of coupling functions~\cite{2018-Sharma.Subramanian.Seshadri.PRD}.

In this work, to avoid the strong-coupling problem and restricted coupling functions, we propose a model that couples the electromagnetic fields with the Riemann tensor. To our knowledge, Riemann tensor coupling has not been discussed in the literature to generate helical fields. The model has three key features: First, it does not require the coupling of the electromagnetic field with the scalar field. Hence, there are no extra degrees of freedom and will not lead to a strong-coupling problem. Second, the conformal invariance is broken due to the coupling to the Riemann tensor. Since the curvature is large in the early Universe, the coupling term will introduce non-trivial corrections to the electromagnetic action. However, at late-times, the new term will not contribute, and the theory is identical to standard electrodynamics\footnote{For instance, at the current epoch, $H_0 \sim 10^{-44} GeV$ and hence, the Riemann coupling term will only contribute in the early Universe and not in the late Universe.}. {The power-spectrum of the fields has slight red-tilt for slow-roll inflation}. Third, as we show explicitly, our model is free from backreaction for a range of scale-factor during inflation. This is different from other models where a specific form of coupling function is chosen to avoid any back-reaction~\cite{2018-Sharma.Subramanian.Seshadri.PRD}. To generate helical fields during inflation, most models in the literature introduce non-minimal coupling to the electromagnetic fields with a scalar field. In the scalar field coupled models, the helical magnetic fields with and without superhorizon correlations are possible. As was shown in Ref.~\cite{2018-Sharma.Subramanian.Seshadri.PRD}, the original Ratra's model can generate superhorizon correlations; however, 
it leads to a strong coupling problem. In Ref.~\cite{2018-Sharma.Subramanian.Seshadri.PRD}, to avoid a strong coupling problem, 
they used a modified form of non-minimal coupling. This model can not generate 
superhorizon correlations and is sensitive to the physics at reheating epoch.
In our model, the primordial magnetic field is not sensitive to physics at the reheating epoch. We show that our model generates the magnetic field of strength $10^{-14} {\rm G}$ over $ \rm{Mpc}$ scale.

In Sec. \eqref{sec:Model}, we introduce the model and discuss its properties. We discuss the classical properties and define the relevant quantities. We also briefly discuss the procedure to quantize these fields in the FRW background. In Sec. \eqref{sec:Helical}, {we explicitly evaluate the helical magnetic field generation in our model and show that the power-spectrum is red-tilted.} This is different compared to the other models in the literature. We also show that the model does not have a backreaction problem for a range of scale-factor. We discuss the implications of the results in Sec. \eqref{sec:conc}.

In this work, we work use $(+,-,-,-)$ signature for the 4-D space-time metric. Greek alphabets denote the 4-dimensional space-time coordinates, and Latin alphabets denote the 3-dimensional spatial coordinates.  A prime stands for a
derivative with respect to conformal time $(\eta)$ and $,i$ denotes a derivative w.r.t space components. We also work in Heaviside-Lorentz units such that $c = k_B = \epsilon_0 = \mu_0 = 1$. The reduced Planck mass is denoted by $M_{\rm P} = (8 \pi G)^{-1/2}$.

%
%%%%%%%%%%  S E C T I O N %%%%%%%%%%%%%%%%%%%%
%
%
\section{The model}
\label{sec:Model}

We consider the following action:
\begin{align}\label{eq:action}
S  = S_{\rm{Grav}} + S_{\phi} + S_{\rm{EM}} + S_{\rm CB}
\end{align}
where $ S_{\rm{Grav}}$ is the Einstein-Hilbert action
\begin{align}\label{eq:EH-action}
S_{\rm Grav} = -\frac{M_{\rm P}^2}{2}\int d^4x \sqrt{-g} \, R \, ,
\end{align}
and $ S_{\phi} $ is the action for the minimally coupled, self-interacting canonically scalar field:
\begin{align}\label{eq:inflation-action}
S_{\phi} = \int d^4x \sqrt{-g} \left[  \frac{1}{2} \partial_{\mu}\phi \partial^{\mu}\phi -  V(\phi) \right].
\end{align}
We assume that the scalar field ($\phi$) dominates the energy density in the early Universe (during inflation) and leads to $60 \, - \, 70$ e-foldings of inflation with $H \simeq 10^{14} {\rm GeV}$. $S_{\rm{EM}}, S_{\rm CB}$ refer to the standard electromagnetic (EM)  and conformal breaking part of the electromagnetic terms, respectively, and given by:
\begin{align}\label{eq:S_EM}
 S_{\rm{EM}} &= -\frac{1}{4} \int d^4x \, \sqrt{-g} \, F_{\mu\nu} F^{\mu\nu}, \hspace{0.5cm}\\  
 \label{eq:S_h}
 S_{\rm{CB}} &= - \frac{\sigma}{M^2} \,\int d^4x \, \sqrt{-g} \, R_{\rho\sigma}\,^{\alpha\beta} F_{\alpha\beta} \, \tilde{F}^{\rho\sigma} = - \frac{\sigma}{M^2} \,\int d^4x \, \sqrt{-g} \, \tilde{R}^{\mu\nu\alpha\beta} F_{\alpha\beta} \, F_{\mu\nu} \, ,
 \end{align}
where $R_{\rho\sigma}\,^{\alpha\beta}$ is the Riemann tensor and its dual is $\tilde{R}^{\mu\nu\alpha\beta} = \frac{1}{2}\epsilon^{\mu\nu\rho\sigma} R_{\rho\sigma}\,^{\alpha\beta}$, $A_{\mu}$ is the four-vector potential of the electromagnetic field, $F_{\mu\nu} = \nabla_{\mu}A_{\nu} - \nabla_{\nu}A_{\mu} $ and $\tilde{F}^{\rho\sigma} = \frac{1}{2} \epsilon^{\mu\nu\rho\sigma}F_{\mu\nu} $ is the dual of $F_{\mu\nu}$. $\epsilon^{\mu\nu\rho\sigma} = \frac{1}{\sqrt{-g}}\, \eta^{\mu\nu\rho\sigma}$ is fully antisymmetric tensor, $\eta^{\mu\nu\rho\sigma}$ is Levi-Civita symbol whose values are $\pm1$ and we set $\eta^{0123} = 1 = - \eta_{0123}$. 

The standard electromagnetic action $S_{\rm{EM}}$ is conformally invariant; however, the presence of Riemann curvature in $S_{\rm CB}$ breaks the conformal invariance. $M$ is the energy scale, which sets the scale for the breaking of conformal invariance. For our discussion below, we assume that
$10^{-3}  \leq (H_{\rm Inf}/M) \leq 1$~\cite{2018-Nakonieczny-JHEP,2016-Goon.Hinterbichler-JHEP,2016-Goon-JHEP,2013-Balakin.etal-CQG} where $H_{\rm Inf}$ is the Hubble scale during inflation which we assumed to be $10^{14}~{\rm GeV}$.   
Due to Riemann coupling, $M$ appears as a time-dependent coupling in the FRW background~\cite{1971-Prasanna-PLA}. To see this, let us evaluate Riemann tensor for the flat FRW background \eqref{eq:FRW}: 
\[
R_{\mu\nu}\,^{\sigma\gamma} \sim \frac{{a^{\prime}}^2}{a^4} ~\rm{or}~ \frac{a^{\prime\prime}}{a^3}
\]
Thus, the coupling function is time-dependent, i. e.,
\begin{align}\label{eq:coupl-I_for_constant_M}
\frac{1}{M_{\rm eff}} \sim \frac{1}{M} \frac{a^{\prime}}{a^2} = \frac{H}{M}\, .
% \propto \frac{1}{\eta^{2\beta + 4}}. 
\end{align} 
 Thus, the effective coupling $1/M_{\rm eff} \sim H/M$ where $H$ is the Hubble parameter in cosmic time. As mentioned above, $H \sim 10^{14} GeV$ during inflation and hence, we assume that $10^{-3}  \leq (H_{\rm Inf}/M) \leq 1$. Let us now evaluate the effective coupling $1/M_{\rm eff}$ at the current epoch where 
$H_0 \approx 10^{-44} \rm{GeV}$ and assuming the parameter $M \approx 10^{17} \rm{GeV}$, we obtain $\frac{H_0}{M} \sim 10^{-61}$. Therefore, the coupling (Riemann tensor) is tiny and the non-minimal coupling term in the electromagnetic action will have significant contribution only in the early universe.

Before proceeding with the analysis, we want to highlight the following salient features of this model compared to the earlier models that introduce non-minimal scalar field coupling in action: First, our model does not require the coupling of the electromagnetic field with the scalar field. Hence, there are no extra degrees of freedom and will not lead to a strong-coupling problem. Second, the conformal invariance is broken due to the coupling to the Riemann tensor. Since the curvature is significant in the early Universe, the coupling term will introduce non-trivial corrections to the electromagnetic action. As mentioned above, at late-times, $S_{\rm CB}$ will not contribute, and the model is identical to standard electrodynamics.  Third, as we show explicitly, our model is free from backreaction for a range of scale-factor during inflation. This is different from other models where a specific form of coupling function is chosen to avoid any back-reaction~\cite{2018-Sharma.Subramanian.Seshadri.PRD}. 

As mentioned earlier, we aim to generate the helical magnetic field during inflation. Hence, the scalar field's energy density dominates over the standard electromagnetic and conformal breaking term in action \eqref{eq:action}.
Since the single-scalar field inflation can not generate vector perturbations, the magnetic field generated will be due to the conformal breaking term $S_{\rm CB}$ in action.
 
The variation of the action (\ref{eq:action}) with respect to gauge field $A_{\mu}$ leads to the following equation:
\begin{align}\label{eq:eom}
\partial_{\mu} \left( \sqrt{-g} \, F^{\mu\nu} + \frac{1}{M^2}  \sqrt{-g} \,  \epsilon^{\alpha\beta\rho\sigma} R_{\rho\sigma}\,^{\mu\nu}   \, F_{\alpha\beta} +  \frac{1}{M^2} \sqrt{-g}   \, \epsilon^{\mu\nu\rho\sigma} R_{\rho\sigma}\,^{\alpha\beta} \, F_{\alpha\beta} \,  \right) = 0
\end{align}
As mentioned above, we will consider a flat  Friedman universe described by the line-element:
\begin{align}\label{eq:FRW}
ds^2 = a^2(\eta) \,(d\eta^2 - \delta_{ij} dx^i dx^j)
\end{align}
where $\eta$ is the conformal time. For $\nu = i$, Eq.~(\ref{eq:eom}) reduces to: 
\begin{align}\label{eq:eom1}
 \partial_{0} \left(F_{0 i} + \frac{2}{M^2}  \,  \eta^{0 i j k}  \frac{a^{\prime\prime} }{a^3} \,F_{j k} \,  \right) 
- \partial_{l} \left( \, F_{l i} - \frac{4}{M^2}  \,  \eta^{0 i l j}  \frac{  a^{\prime\prime} }{a^3}  F_{j0} \,\right)   = 0  
\end{align}
where we have substituted the following components of Riemann tensor: 
\[
R_{i j}\,^{k l} =  \frac{    {a^{\prime} }^2}{a^4} \, \left( \delta^l_i \,  \delta^k_j  - \delta^k_i \,  \delta^l_j  \right),  R_{0 i}\,^{0 j} =  \left( \frac{ {a^{\prime}}^2}{a^4} -\frac{a^{\prime\prime}}{a^3} \right) \, \delta^j_i \, .
\]
In the Coulomb gauge ($A^{0} = 0, \partial_iA^i = 0$), and using $\eta^{0 i j l} = \epsilon_{i j l}$ (where $\epsilon_{i j l}$ is the Levi-Civita symbol in 3D Euclidean space), the above equation motion (of $A^i$) leads to the following evolution equation:
\begin{align}\label{eq:equation_of_motion}
A_i^{\prime\prime} + \frac{4 \, \epsilon_{i j l}}{M^2} \, \left( \frac{a^{\prime\prime\prime}}{a^3} - 3\frac{a^{\prime\prime} a^{\prime} }{a^4} \right) \partial_j A_l 
- \partial_j \partial_j A_i = 0
\end{align}
Note that the above equation differs from other models in the literature by an overall factor in the term with $\epsilon_{i j l} \partial_j A_l$, especially, the third derivative of the scale factor. Thus, the model can lead to different evolution of the fluctuations in comparison to non-minimally coupled scalar field models.

In this work, we consider two inflationary scenarios --- power-law inflation and the slow-roll inflation --- to evaluate the power-spectrum of the electromagnetic fluctuations. For power law inflation the scale factor (in cosmic time) is given by  $a(t) = a_0 t^p$ where $p>1$ and $a_0$ is arbitrary constant. In the conformal time, the scale factor is~\cite{2004-Shankaranarayanan.Sriramkumar-PRD}:
 \begin{align}\label{eq:powerLaw}
a(\eta) =  \left( - \frac{\eta}{\eta_0} \right)^{(\beta+1)}
\end{align}
where $\eta_0$ is an arbitrary constant and denotes the scale of inflation. During inflation, $\eta \in (-\infty, 0)$.  $\beta$ and $\eta_0$ are given by:
\begin{align}\label{eq:mathcalH}
\beta = - \left( \frac{2p - 1}{p - 1} \right)  \hspace{0.5cm} \text{and} \hspace{0.5cm} \eta_0 = \left[ (p - 1)a_0^{1/p}  \right]^{-1}.
\end{align}
Note that $\beta \leq -2$ and $\beta = -2$ corresponds to the de Sitter. The Hubble parameter ($\mathcal{H} \equiv {a^{\prime}(\eta) }/{a(\eta)}$) is given by:
 \begin{align}\label{eq:H-eta_relation}
 \mathcal{H} = \frac{a^{\prime} }{a } = \frac{\beta + 1}{\eta} \implies \eta = \frac{\beta + 1}{\mathcal{H}}
 \end{align}
Slow-roll inflation is a generic inflationary paradigm that leads to an accelerated expansion independent of a particular model (or potential). In this case, we have 
\begin{equation}
\beta \approx -2-\epsilon, \quad \mathcal{H} \approx - \frac{1 + \epsilon}{\eta}   
\end{equation}
where $\epsilon $ is the slow roll parameter. 

\subsection{Physical quantities of interest}

Although, the four-vector potential $A^{\mu}$ provides the covariant
description of the electromagnetic processes, to compare with observations, we need to decompose the physical quantities in terms of the electric and magnetic fields, that are intrinsically frame-dependent. Hence, it is always useful to define a comoving observer with velocity $u^{\mu} = \left( 1/a(\eta),0, 0, 0 \right)$ satisfying $u_{\mu} u^{\mu} = 1$. The electric and magnetic field four-vector for this observer is given by projecting the EM field tensor with $u^{\mu}$ as
\begin{align}
E_{\mu} = u^{\alpha} F_{\alpha\mu}, \hspace{.5cm}
B_{\mu} = \frac{1}{2} \,u^{\gamma} \, F^{\alpha\beta} \epsilon_{\gamma\alpha\beta\mu}   = u^{\alpha}\tilde{F}_{\alpha\mu} .
\end{align}
Note that the electric and magnetic field four-vectors are both three-vector fields in a sense that they are orthogonal to the comoving observer, i.e., $E_{\mu} u^{\mu} = 0 = B_{\mu} u^{\mu}$, and we have:
\begin{align}
E_{\mu} =  a(\eta) \left( 0, \textbf{E} \right), \quad
B_{\mu}  =  a(\eta) \left( 0, \textbf{B} \right) 
\end{align}
where 
\[
 \textbf{E} =   \frac{A_i^{\prime}}{a^2(\eta)},  \quad  \textbf{B} = -\frac{1 }{a^2(\eta)}\,  \epsilon_{ijk}  \, \partial_i A_j \, . 
 \]
Electromagnetic energy densities are defined as
\begin{align}
\rho_B \equiv - \frac{1}{2} B_{\mu} B^{\mu} = \frac{1}{2}  \textbf{B} \cdot \textbf{B}, \qquad
\rho_E \equiv - \frac{1}{2} E_{\mu} E^{\mu} = \frac{1}{2}  \textbf{E} \cdot \textbf{E}.
\end{align}
and the magnetic helicity density is
\begin{align}
\rho_h \equiv - A_{\mu} B^{\mu}.
\end{align}
We will evaluate these quantities for the quantum fluctuations generated during inflation.

\subsection{Quantization in the Helicity basis}
In this section, we briefly discuss the evolution of the quantum fluctuations of the electromagnetic field in the helicity basis~~\cite{2018-Sharma.Subramanian.Seshadri.PRD}.  Decomposition of the vector potential in Fourier space, we have:
\begin{align}\label{eq:FourierT}
A^{i}(\vec{x}, \eta) =  \int \frac{d^3 k}{(2\pi)^3} \sum_{\lambda = 1,2} \varepsilon^i_{\lambda} \left[ A_{\lambda}(k,\eta) b_{\lambda}(\vec{k}) e^{ik\cdot x}  
+ A^*_{\lambda}(k,\eta)  b^{\dagger}_{\lambda}(\vec{k}) e^{- ik\cdot x} \right]
\end{align}
where $b(\textbf{k})$ and $b^{\dagger}(\textbf{k})$ are the annihilation and creation operators respectively for a given comoving mode $\textbf{k}$, and $\varepsilon_{\lambda}^i$ is the orthogonal basis vector which in right-handed coordinate system~\cite{2018-Sharma.Subramanian.Seshadri.PRD} is given by
\begin{align}\label{eq:basisVector}
\varepsilon^{\mu} = \left( \frac{1}{a}, \textbf{0} \right), \,\,\,\, \varepsilon^{\mu} = \left( 0, \frac{ \hat{\varepsilon}^i_{\lambda} }{a} \right), \,\,\,\, \varepsilon^{\mu}_3 = \left(  0, \frac{\hat{\textbf{k}}}{a} \right) \quad  \text{for} \quad \lambda = 1, 2 \, ,
\end{align}
3-vectors $\hat{\varepsilon}^i_{\lambda}$ are unit vectors orthogonal to $\hat{\textbf{k}}$ and to each other. Substituting Eq.~(\ref{eq:basisVector}) in 
Eq.~(\ref{eq:FourierT} ) and defining the new variable 
$\bar{A}_{\lambda} = a(\eta) \,  A_{\lambda}(k,\eta)$, we have:
\begin{align}\label{eq:Fdecomposition}
A_{i}(\textbf{x}, \eta) = \int \frac{d^3 k}{(2\pi)^3} \sum_{\lambda = 1,2} \,\hat{\varepsilon}_{i \lambda} \left[ \bar{A}_{\lambda} b_{\lambda}(\textbf{k}) e^{ i \textbf{k} \cdot \textbf{x} }  
+ \bar{A}^*_{\lambda}  b^{\dagger}_{\lambda}(\vec{k}) e^{- i \textbf{k} \cdot \textbf{x} } \right] \, .
\end{align}
Substituting  Eq.~\eqref{eq:Fdecomposition} in Eq.~\eqref{eq:equation_of_motion}, we get:
\begin{align}\label{eq:EOM_fourier_space}
\sum_{\lambda = 1,2}b_{\lambda} \left[  \hat{\varepsilon}_{i \lambda}  \bar{A}_{\lambda}^{\prime\prime} + \frac{4i}{M^2} \epsilon_{i j l} k_j \hat{\varepsilon}_{l \, \lambda} \bar{A}_{\lambda} \, \left( \frac{a^{\prime\prime\prime} }{a^3} - 3\frac{ a^{\prime\prime}  a^{\prime}  }{a^4} \right) +  k^2 \hat{\varepsilon}_{i \lambda} \bar{A}_{\lambda}\right] = 0 
\end{align}
where we have used $\partial_j \partial_j = -k^2$. 

Since the action \eqref{eq:action} contains parity breaking term (helicity term), it is always useful to work in the helicity basis. The helicity basis vectors $\varepsilon_+$ and $\varepsilon_-$ corresponding to $h = +1$ and $h = -1$ are defined as
\begin{align}\label{eq5:helicity_basis}
\varepsilon_{\pm} = \frac{1}{\sqrt{2}} \left(   \hat{\varepsilon}_1 \pm i  \hat{\varepsilon}_2  \right).
\end{align}
Assuming that the wave propagates in the $z-$direction, the vector potential in the helicity basis is given by:
%the case where the wave is propagating along the $\varepsilon_3$ direction, in %coulomb gauge (radiation gauge), the vector potential takes the form
%
\begin{align}\label{eq5:decomp_A_helicity}
\bar{\textbf{A}} = \bar{A}_1 \hat{\varepsilon}_1 + \bar{A}_2  \hat{\varepsilon}_2 = A_+ \varepsilon_+ + A_- \varepsilon_-
\end{align}
where $A_+$($A_-$) refer to the vector potential with positive (negative) helicity. 
The ground state in the helicity basis is defined  as
\begin{align}\label{eq:GS_helicity}
b_h(\textbf{k}) | 0 \rangle = 0 
\end{align}
and commutation relation are:
\begin{align}\label{eq:comm-b_h}
\left[ b_h(\textbf{k}), b^{\dagger}_{h^{\prime}}(\textbf{q})  \right] &= \left( 2\pi \right)^3 \, \delta^3(\textbf{k} - \textbf{q}) \, \delta_{h h^{\prime}} \\
\left[ b_h(\textbf{k}), b_{h^{\prime}}(\textbf{q})  \right] &= 0 = \left[ b^{\dagger}_h(\textbf{k}), b^{\dagger}_{h^{\prime}}(\textbf{q})  \right] \, .
\end{align}

Rewriting \eqref{eq:EOM_fourier_space} in the Helicity basis and replacing
$\epsilon_{i j l} \partial_j A_l \longrightarrow  -k \sum_{h = \pm 1} h A_h \varepsilon_{h}$, we have:
\begin{align}\label{eq:eom_helicity}
A_h^{\prime\prime} + \left[  k^2 - \frac{4kh}{M^2} \, 
\Gamma(\eta)  \right] A_h= 0 \, ,
\end{align}
where,
\begin{equation}
\label{def:Gamma}
  \Gamma(\eta) = \frac{a^{\prime\prime\prime}}{a^3} - 3\frac{a^{\prime\prime} a^{\prime} }{a^4} = \frac{1}{a^2} \left(\mathcal{H}'' - 2 \mathcal{H}^3\right) \, .  
\end{equation}
We would like to stress the following points regarding the above expression: First, unlike the scalar or tensor perturbations, the mode functions contain third-order derivatives of the scale factor. This implies that the spectrum of perturbations may be different in our model. Second, since the perturbations equations contain second-order derivatives of $\mathcal{H}$, the helicity modes will be different for inflation and bounce models~\cite{2020-Nandi-PLB}. Third, since $h$ takes two values, $A_h$ evolves differently for the two modes leading to non-zero helicity. 

The EM energy densities of the ground state with respect to the comoving observer are:
\begin{align}\label{eq:rhoB}
& \rho_B\left(\eta, k \right) \equiv -\frac{1}{2}\langle 0 | B_i B^i | 0 \rangle = \int \frac{dk}{k} \, \frac{d\rho_B}{d\rm{ln}k} = \int \frac{dk}{k} \frac{1}{\left( 2\pi \right)^2} \frac{k^5}{  a^4} \left( \,\, \left| A_+\left(\eta, k \right) \right|^2 + \left| A_- \left(\eta, k \right) \, \, \right|^2  \right) \\
\label{eq:rhoE}
& \rho_E\left(\eta, k \right) \equiv -\frac{1}{2}\langle 0 | E_i E^i | 0 \rangle = \int \frac{dk}{k} \, \frac{d\rho_E}{d\rm{ln}k} = \int \frac{dk}{k} \frac{1}{\left( 2\pi \right)^2}  \frac{k^3}{ a^4} \left( \,\,  \left| A^{\prime}_+\left(\eta, k \right) \right|^2 + \left| A^{\prime}_-\left(\eta, k \right) \right|^2 \,\, \right)
\end{align}
and the ground state helicity density as
\begin{align}\label{eq:rhoh}
\rho_h \left(\eta, k \right) \equiv -\langle 0 | A_i B^i | 0 \rangle = \int \frac{dk}{k} \, \frac{d\rho_h}{d\rm{ln}k} = \int \frac{dk}{k} \frac{1}{2\pi^2} \frac{k^4}{ a^3} \left( \,\, \left| A_+\left(\eta, k \right) \right|^2 - \left| A_-\left(\eta, k \right) \right|^2 \,\,  \right).
\end{align}
where $d\rho_{\Upsilon}/d (\ln k)$ for $\Upsilon \in \{ E,B,h\}$ is the spectral energy contained in logarithmic interval in $k-$space. Note that the helicity density is the difference between the two helicity spectrum. Hence, it is possible to maximize the magnetic helicity density, if one helicity is enhanced and the other helicity is suppressed~\cite{2013-Durrer.Neronov-Arxiv}. For most of the calculation, we will keep both the terms and evaluate the energy density for both helicity modes. 

%
%%%%%%%   S E C T I O N %%%%%%%%%%%%%%%%
%

\section{Helical magnetic field generation}
\label{sec:Helical}

In this section, we explicitly calculate the power-spectrum and energy densities for our model in power-law inflation. We obtain the power-spectrum in the slow-roll limit. 

Substituting the power-law scale factor (\ref{eq:powerLaw}) in Eq.(\ref{eq:eom_helicity}) leads to:
\begin{align}\label{eq:arbitrayBeta_eta}
{A_h^{\prime\prime} + \left[ k^2 - \frac{8kh}{M^2} 
\frac{ \beta (\beta+1) ( \beta + 2)}{\eta_0^3} 
\left( \frac{-\eta_0}{\eta} \right)^{(2 \beta+5)} \right]  \, A_h = 0 }
\end{align}
As expected, for de-sitter case ($\beta = -2$), the helicity term $(\Gamma(\eta)$) vanishes. This is consistent with the fact that the de Sitter symmetry will not be preserved in the presence of helicity terms. However, it will be non-zero for the approximately de-sitter universe i.e., $\beta = -2-\epsilon$. 

For the power-law inflation model, the scalar and tensor perturbations can be evaluated exactly. However, as can be seen, it is not possible to obtain an exact expression. To obtain the solution, we consider two regions. In Region I (sub-horizon limit), the wavelength of the mode is smaller than the Hubble radius, i. e., $H \ll k$. In this region, we can neglect $\Gamma(\eta)$ in Eq. \eqref{eq:arbitrayBeta_eta}.  In Region II (super-Horizon scales), the mode is outside the Hubble radius i. e., $k \ll H$. In this region, we can neglect $k^2$ in Eq. \eqref{eq:arbitrayBeta_eta}. The constants are fixed by matching 
$A_h$ and $A_h'$ at the transition time between regions I and II
at $\eta_*$. While evaluating the mode-functions is trivial in Region I, it is highly non-trivial in Region II. In the rest of this section, we obtain the mode functions and calculate the power-spectrum.

In Region I ($\left| - k \eta \right| \gg 1$), Eq.~(\ref{eq:arbitrayBeta_eta}) simplifies to:
\begin{align}\label{eq:sub-horizon}
A_h^{\prime\prime} + k^2 A_h \approx 0 
\end{align}
and assuming that the quantum field is in the vacuum state at asymptotic past 
(Bunch-Davies vacuum state), we have:
\begin{align}
A_h = \frac{1}{\sqrt{k}} e^{-ik\eta}
\end{align}
In Region II ($\left| - k \eta \right| \ll 1$),  Eq.~(\ref{eq:arbitrayBeta_eta}) becomes:
\begin{align}\label{eq:sup_mode_eq_alpha}
{A_h^{\prime\prime} + h  k \frac{\varsigma^2}{\eta^2} 
\left( \frac{-\eta_0}{ \eta} \right)^{2 \alpha }   \, A_h = 0 }
\end{align}
where  
\begin{equation}
\label{def:varsigma}
{\varsigma^2 \equiv -\frac{1}{M^2 \, \eta_0} (2\alpha - 3)(2\alpha - 1)(2\alpha + 1)\, ,
\quad \alpha = \beta + \frac{3}{2} }
\end{equation}
$\alpha$ makes the expressions look tidier! Note that $\alpha = -\frac{1}{2}$ corresponds to de-sitter and $ \alpha \leq - \frac{1}{2} $.

As mentioned above, unlike in the scalar and tensor perturbations during inflation, the above equation is not exactly solvable. To do this, we introduce a new dimensionless variable $\tau$, and is defined as:
\begin{align}\label{eq:tau_to_eta}
\tau = \left( -\frac{\eta_0}{ \eta} \right)^{\alpha} \quad \implies  \quad \eta = - \frac{\eta_0}{\tau^{\frac{1}{\alpha}}} \, .
\end{align}
$\tau$ and $\eta$ are monotonically related while $\tau$ is a positive definite quantity ($0 < \tau < \infty$). [At the start of inflation, $\tau$ is large and vanishes 
at the end of inflation.] In terms of $\tau$, the scale factor for the power-law inflation \eqref{eq:powerLaw} is 
\[
a (\tau)= \left( \frac{1}{\tau} \right)^{1 - \frac{1}{2\alpha}} \, . 
\]
Rewriting Eq. \eqref{eq:sup_mode_eq_alpha} in terms of $\tau$  \eqref{eq:tau_to_eta}, we have:
\begin{align}\label{eq:sup_mode_alpha_tau-varsigma}
{   \alpha^2 \frac{d^2 A_h}{d\tau^2} + \frac{\alpha(\alpha+1)}{\tau} \frac{d A_h}{d \tau} + h \, k \, \varsigma^2 A_h  = 0     }
\end{align}
The above equation is a Bessel differential equation, and it has a complete solution as
\begin{subequations}
\begin{align}\label{eq:sup_mode_h+}
A_{+}(\tau,k) &= \tau^{- \frac{1}{2\alpha} } \, J_{  \frac{1}{2 \alpha}} \left( \frac{\varsigma \sqrt{k} }{\alpha} \tau \, \right)  C_1+ \tau^{- \frac{1}{2\alpha} } \, Y_{ \frac{1}{2 \alpha} }  \left( \frac{ \varsigma \sqrt{k} }{\alpha}\tau \right)  C_2
\\
\label{eq:sup_mode_h-}
A_{-}(\tau,k) &=  \tau^{- \frac{1}{2\alpha} } \, J_{  \frac{1}{2 \alpha}} \left(  -i \frac{ \varsigma \, \sqrt{k} }{\alpha} \tau  \right)  C_3+ \tau^{- \frac{1}{2\alpha} }  \, Y_{  \frac{1}{2 \alpha} } \left(  - i \, \frac{ \varsigma\, \sqrt{k} }{\alpha} \tau  \right)  C_4
\end{align}
\end{subequations}
where $C_1, C_2, C_3, C_4$ are arbitrary constants of dimension $L^{1/2}$. 
As mentioned above, for the two helicity modes, we fix the constants $C_1, C_2$ ($C_3, C_4)$ by matching $A_h$ and $A_h'$ at the transition time between between regions I and II at $k_* \sim \eta_*^{-1}$ where $*$ refers to the 
quantities evaluated at the horizon-exit.

Although the analysis can be done for any general value of $\alpha$, to keep the 
calculations tractable, we obtain the constants for $\alpha = -1$.  There are two reasons for this choice: First, in this special case, $\tau \propto \eta$ and the 
super-horizon modes can be written in terms of $\eta$ using the linear relation. 
Second, the constants $C_1, C_2, C_3, C_4$ have a weak dependence of $\alpha$ 
and, hence, finding the value for a given value of $\alpha$ will be accurate within a order. Thus, matching he solutions and the derivatives at the horizon-exit, we get:
\begin{align}\label{eq:Coefficients}
C_1 &= -e^i \,  \sqrt{ \frac{\pi \eta_0}{ 2} } \left( \frac{1}{\sqrt{\Theta}} \rm{sin}\Theta 
 + i \sqrt{ \Theta }  \,  \rm{cos} \Theta   \right), \,\,\,\,
C_2 = -i \, e^i \,  \sqrt{ \frac{\pi \eta_0}{ 2} } \left( \frac{1}{\sqrt{\Theta}} \rm{cos}\Theta 
 - i \sqrt{ \Theta }  \,  \rm{sin} \Theta   \right) \\
C_3 &= e^i \,  \sqrt{ \frac{\pi \eta_0}{ 2 } } \left( \frac{1}{\sqrt{i \Theta}} \rm{sinh}\Theta 
 +  \sqrt{ i \Theta }  \,  \rm{cosh} \Theta   \right), \,\,\,\,
C_4 = -i \, e^i \,  \sqrt{ \frac{\pi \eta_0}{ 2 } } \left( \frac{1}{\sqrt{i \Theta}} \rm{cosh} \Theta
 +  \sqrt{ i \Theta }  \,  \rm{sinh} \Theta   \right). \nonumber
\end{align}
where $\Theta = \sqrt{ \frac{15 \eta_*}{M^2 \eta_0^3} }$ is the dimensionless constant.

\begin{figure}
\centering
\subfigure[]{%
\label{fig:first}%
\includegraphics[height=2in]{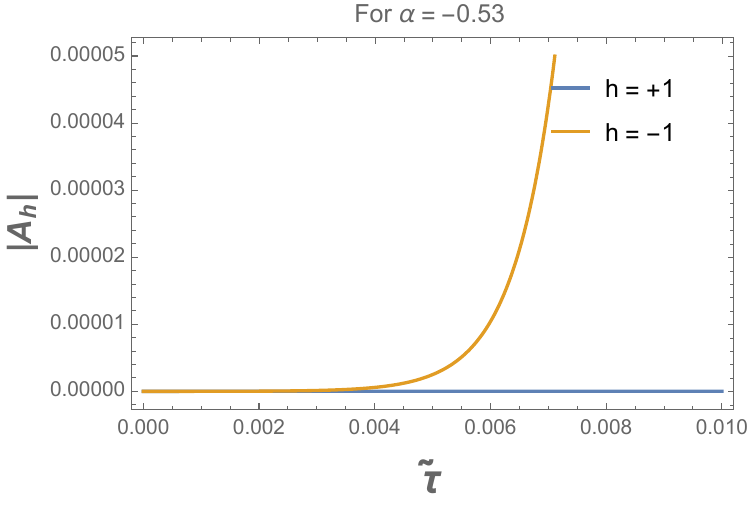}}%
\qquad
\subfigure[]{%
\label{fig:second}%
\includegraphics[height=2in]{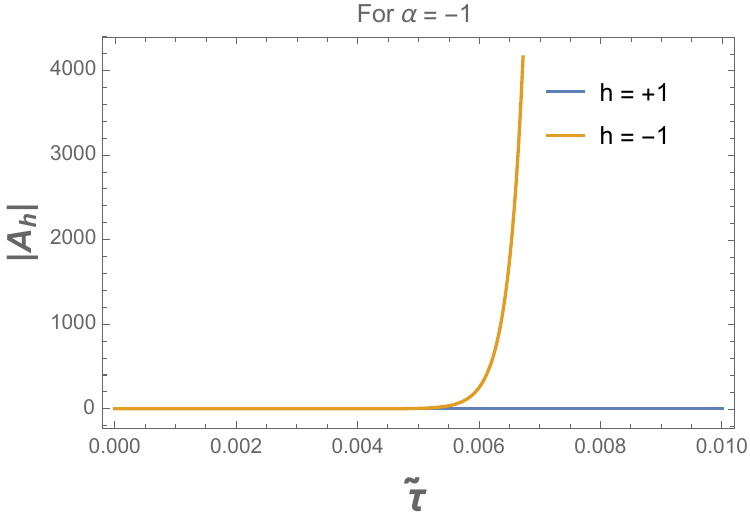}}%
\caption{Figure showing the behaviour of positive and negative helicity mode for \ref{fig:first} $\alpha = -0.53$ and \ref{fig:second} $\alpha = -1$. $\tilde{\tau} = 10^{-\frac{63}{2}} \,\tau $ and the vertical axis is in $\rm{GeV}^{-1/2}$. $\alpha = -0.53$ corresponds to approximate de-Sitter.}
\label{fig:helicitymode}
\end{figure}

To obtain the dominating helicity mode during inflation, we need to obtain the 
values of the coefficients $C_1, C_2, C_3, C_4$. To obtain these values, we 
take: $\mathcal{H} \sim {\eta_0}^{-1} \sim 10^{14} \rm{GeV} \sim {10^{52}} \rm{Mpc}^{-1}$, and $M \sim 10^{17} \rm{GeV}$~\cite{2004-Shankaranarayanan.Sriramkumar-PRD}. This gives $\Theta \sim 10^{-3} $ which is small value. Approximating trigonometric functions in Eq.~(\ref{eq:Coefficients}), we obtain
\begin{align}\label{eq:Coeff-value}
\left| C_1  \right| \approx \left| C_3  \right| \approx 10^{-17/2} \rm{GeV}^{-\frac{1}{2}}, \hspace{0.5 cm} \text{and} \,\, \,\,\,
\left| C_2  \right| \approx \left| C_4  \right| \approx 10^{-11/2} \rm{GeV}^{-\frac{1}{2}}.
\end{align}
Using these values in Eqs.~(\ref{eq:sup_mode_h+}, \ref{eq:sup_mode_h-}), 
we have plotted the two modes for $\alpha = -0.53$ and $\alpha = -1$ in 
\ref{fig:helicitymode}. The plots show that the positive helicity modes are growing compared to the negative helicity modes. Specifically, from Fig.\ref{fig:second}, we see that negative helicity mode is decaying. (For $\alpha = -1$ we have $\tau \propto -\eta$ which means negative mode is decaying from $-\infty$ to zero in conformal time). Hence, we can set $\left|  A_{-}(\tau,k) \right| = 0$. The helicity density \eqref{eq:rhoh} is the difference between the two helicity spectrum, and maximum helicity is achieved if one helicity is enhanced compared to other.  In our case, the negative helicity mode is negligible and, has been set to zero.

Using the series expansion of the Bessel functions (see appendix \ref{app:Series-exp} for details), in the leading order, positive helicity mode (\ref{eq:sup_mode_h+}), takes the following form: 
%
%\begin{subequations}
\begin{align}\label{eq:A+Series}
A_+(\tau,k) &= C \, k^{\frac{1}{4\alpha}} 
  - C_2 \frac{\mathcal{F}^{-1} }{\pi} 
  \Gamma \left( \frac{1}{2\alpha}  \right) \,k^{-\frac{1}{4\alpha}} \tau^{ - \frac{1}{\alpha} }  
  %
%  \label{eq:A-Series}
%A_-(\tau,k) &=\tilde{C} \, k^{\frac{1}{4\alpha}} 
%-C_4 \frac{\tilde{\mathcal{F}}^{-1} }{\pi} 
%  \Gamma \left( \frac{1}{2\alpha}  \right) \, k^{-\frac{1}{4\alpha}} \tau^{ -\frac{1}%{\alpha} } .
\end{align}
%\end{subequations}
%
where $C, \mathcal{F}$ are constants defined in Appendix \eqref{app:Series-exp}. The value of these constants are evaluated using (\ref{eq:Coeff-value}). Note that $\varsigma \approx 10^{-10} \rm{GeV}^{-1/2}$ \, the floor function $F(\tau) = 1$. 
Fig.~(\ref{fig:Plot}) contains the plot of Floor function for the range of parameters used in evaluated  (\ref{eq:Coeff-value}). Also, $\left| \mathcal{F} \right|  \sim 10^{-\frac{5}{\alpha}} \,\, \rm{GeV}^{-1/4\alpha}, \text{and} \,
\left| C \right| \sim 10^{-\frac{5}{\alpha} - \frac{11}{2}}$ $\rm{GeV}^{-\frac{1}{4\alpha} - \frac{1}{2}}$.
%
%
%
%%%%%%%%%   S U B - S E C T I O N %%%%%%%%%%%%%%%%%
%
%
\subsection{Energy densities, power spectrum, and backreaction}

Substituting Eq.~(\ref{eq:A+Series}) in the relation (\ref{eq:rhoB}), 
the spectral magnetic energy density is given by:
\begin{align}\label{eq:spetralB1}
\frac{d\rho_B}{d\rm{ln}k} =  \frac{1}{\left( 2\pi \right)^2} \frac{k^5}{  a^4(\eta(\tau))}
 \left(  \,\,\,\,   \left|   C \right|^2 \, k^{\frac{1}{2\alpha}} 
  + \left| C_2 \frac{\mathcal{F}^{-1} }{\pi} 
  \Gamma \left( \frac{1}{2\alpha}  \right)  \right|^2   \,
  \left( k \,   \tau^4 \right)^{ - \frac{1}{2\alpha} }     \right).  
\end{align}
{Using  the expression: $\tau_* =   \left(  -\frac{2 \, \eta_0 k_*}{2\alpha - 1}  \right)^{\alpha}$ at the exit of inflation, spectral energy density at horizon exit is given by:}
\begin{align}\label{eq:powerSpetrum}
 \left. \frac{d\rho_B}{d\rm{ln}k}  \right|_{k_* \sim \mathcal{H} } =  \frac{(-\eta_0)^{4\alpha - 2 }}{\left( 2\pi \right)^2 }   \left[ \, \left| C \right|^2 \,  k_*^{3 + 4\alpha +\frac{1}{2\alpha}}  +  \left| C_2 \frac{\mathcal{F}^{-1} }{\pi}  \Gamma \left( \frac{1}{2\alpha}  \right) \right|^2 \,
    \frac{ (2\alpha - 1)^2 }{ 4 \eta_0^2} \, k_*^{1 + 4\alpha - \frac{1}{2\alpha}}     \right]        
\end{align}  
 Let us understand the properties of the spectral energy density obtained above: First, 
it has two branches. The first branch (setting $C_2 = 0$) has
scale-invariant spectrum  for $\alpha = -\frac{1}{2}, -\frac{1}{4}$.  Similarly, 
the second branch (setting $C = 0$) has scale invariant spectrum  for $\alpha = -\frac{1}{2}, \frac{1}{4}$. Note that the physically allowed values of $\alpha \leq -1/2$. Hence, $\alpha = \pm 1/4$ is ruled out. Thus, the two branches has scale-invariant power-specrum for exact de Sitter  ($\alpha = -\frac{1}{2}$). Second, 
for slow-roll type of inflation $\alpha = -\frac{1}{2} - \epsilon$, the two branches scale differently ---  $k_*^{-2\epsilon}$ (first branch) and $k_*^{-6\epsilon}$ (second branch). Since $\epsilon$ is positive, this implies that our model produces more power on the large scales. Thus, our model predicts a red spectrum for the helical modes for slow roll inflation. In the next subsection, we compare the results of our model with other models. 

Thus, the magnetic energy density can be obtained as: 
\begin{align}\label{eq:spetralB}
 \rho_B =  \int_{ \frac{\mathcal{H}}{100}}^{\mathcal{H}} \frac{dk}{k} \frac{d\rho_B}{d\rm{ln}k} =  \frac{1}{\left( 2\pi \right)^2 a^4(\eta(\tau)) }   \left[  \left| C\right|^2 \,  \frac{\mathcal{H}^{5 +\frac{1}{2\alpha}} }{ 5 +\frac{1}{2\alpha}  }  +  \left| C_2 \frac{\mathcal{F}^{-1} }{\pi}  \Gamma \left( \frac{1}{2\alpha}  \right) \right|^2 \,
     \tau^{ - \frac{2}{\alpha} }  \frac{\mathcal{H}^{5 - \frac{1}{2\alpha}} }{ 5 -\frac{1}{2\alpha}  }     \right]   
\end{align}
where we have used $\mathcal{H}^{5 + \frac{1}{2\alpha}} - \left(\frac{\mathcal{H}}{100}\right)^{5 + \frac{1}{2\alpha}} 
%= \mathcal{H}^{5 + \frac{1}{2\alpha}} \left( 1 - \frac{1}{100^{5 + \frac{1}{2\alpha}}} %\right) 
\approx \mathcal{H}^{5 + \frac{1}{2\alpha}} $.
 Total magnetic energy density at the exit of inflation can be obtained as   
\begin{align}\label{eq:powerSpetrumB}
\left. \rho_B \right|_{k_* \sim \mathcal{H} } =  \frac{(-\eta_0)^{4\alpha - 2 }}{\left( 2\pi \right)^2 }   \left[ \, \left| C \right|^2 \, \frac{ k_*^{3 + 4\alpha +\frac{1}{2\alpha}} }{ 5 +\frac{1}{2\alpha}  } +  \left| C_2 \frac{\mathcal{F}^{-1} }{\pi}  \Gamma \left( \frac{1}{2\alpha}  \right) \right|^2 \,
    \frac{ (2\alpha - 1)^2 }{ 4 \eta_0^2} \, \frac{ k_*^{1 + 4\alpha - \frac{1}{2\alpha}} }{ 5 -\frac{1}{2\alpha}  }     \right]    
\end{align} 
Using the fact that at super-horizon scales, we can approximate $\partial_{\eta} \sim \mathcal{H}$ (see, for instance, Ref.~\cite{2011-Durrer.Hollenstein.Jain-JCAP}), the total electric energy density at horizon exit is given by: 
\begin{align}\label{eq:powerSpetrumE}
\left. \rho_E \right|_{k_* \sim \mathcal{H} } &=  \frac{(-\eta_0)^{4\alpha - 2 }}{\left( 2\pi \right)^2 }   \left[ \,  \left| C\right|^2 \,  \frac{ k_*^{3 + 4\alpha +\frac{1}{2\alpha}} }{ 3 +\frac{1}{2\alpha}  } + \left| C_2 \frac{\mathcal{F}^{-1} }{\pi}  \Gamma \left( \frac{1}{2\alpha}  \right) \right|^2 \,
    \frac{ (2\alpha - 1)^2 }{ 4 \eta_0^2} \, \frac{ k_*^{1 + 4\alpha - \frac{1}{2\alpha}} }{ 3 -\frac{1}{2\alpha}  }     \right].
\end{align}
Thus, using the above two expressions, the total energy density at horizon exit is given by 
 \begin{align}\label{eq:totalenergy-EB}
\left. ( \rho_B + \rho_E) \right|_{k_* \sim \mathcal{H} } &= \rho_{T}^{(1)} + \rho_{T}^{(2)} 
\end{align}
where
\begin{align}
 \rho_{T}^{(1)} &= \frac{(-\eta_0)^{4\alpha - 2 }}{\left( 2\pi \right)^2 } \, \left| C \right|^2 \frac{ 4\alpha (8\alpha + 1) }{ (10\alpha + 1)(6\alpha + 1)  } k_*^{3 + \alpha +\frac{1}{2\alpha}} \\
 \rho_{T}^{(2)} & =   \frac{(-\eta_0)^{4\alpha - 4 }}{\left( 2\pi \right)^2 }   \left| C_2 \right|^2  \, \, \left| \frac{\mathcal{F}^{-1} }{\pi}  \Gamma \left( \frac{1}{2\alpha}  \right) \right|^2  \frac{ 4\alpha (8\alpha - 1) (2\alpha - 1)^2  }{(10\alpha - 1)(6\alpha - 1)  }  k_*^{1 + 4\alpha - \frac{1}{2\alpha}}  
\end{align}
Since the power-spectrum is red-titled, the power in the long wavelengths is more than in the short wavelengths. Thus, there is a possibility that these helical modes can backreact on the metric. Since the effect is cumulative, all fluctuation modes contribute to the change in the background geometry. Consequently, the backreaction effect can be large, even if the amplitude of the fluctuation spectrum is small. To identify whether these modes lead to backreaction on the metric, we define $R$, which is the ratio of the total energy density of the  fluctuations and 
background energy density during inflation~\cite{2020-Talebian.etal-arXiv}: 
\begin{align}
R = \frac{ \left. ( \rho_B + \rho_E) \right|_{k_* \sim \mathcal{H} } }{6 M_P^2H^2} \, .
\end{align}
Using $M_P = 10^{19}\rm{GeV} $ and $H = 10^{15}~\rm{GeV} $, the background energy density during inflation ($M_P^2H^2$) is $10^{68}~\rm{GeV}^4$. The table below contains estimates of the total energy density at the horizon exit for different values of $\alpha$. 
\begin{table}[h]
\centering
\begin{tabular}{|c|l|l|l|c| }
	\hline
	$\alpha$  &  $\rho^{(1)}_{T}$ (in $\rm{GeV}^4$)  & $\rho^{(2)}_{T}$ (in $\rm{GeV}^4$)& Total (in $\rm{GeV}^4$) &~~~$R$~~~~ \\ \hline
 $-\frac{1}{2} - \epsilon$ & $ \sim 10^{64} $  &  $ \sim 10^{52}$ & $ \sim 10^{64} $  & $\sim 10^{-4}$  \\ \hline
 $-\frac{3}{4} $ & $  \sim 10^{62}$ &  $\sim 10^{54}$ & $ \sim 10^{62} $ 
 & $\sim  10^{-6}$  \\ \hline
 $-1 $ & $  \sim 10^{61}$ &  $\sim 10^{55}$ & $ \sim 10^{61} $  & $\sim  10^{-7}$ 
 \\ \hline
$-3 $ & $  \sim 10^{59}$ &  $\sim 10^{57}$  & $ \sim 10^{59} $ & $\sim 10^{-9}$  
\\ \hline
\end{tabular}
\caption{The total energy density at the exit of inflation for different values of $\alpha$. To estimate  $\rho^{(1)}_{T}$ and $\rho^{(2)}_{T}$, we 
take: $\mathcal{H} \sim {\eta_0}^{-1} \sim 10^{14}~\rm{GeV} \sim {10^{52}}~ \rm{Mpc}^{-1}$, and $M \sim 10^{17}~\rm{GeV}$.}
\end{table}
We see from the above table that for varied values of $\alpha$, $R \ll 1$, implying that the backreaction of the helical modes on the background metric during inflation is negligible. The ratio $R$ is maximum ($10^{-4}$) for slow-roll 
inflation. Thus, while our model produces helical magnetic fields with more power at large length-scales, the backreaction of these on the metric is negligible, and these modes do not stop inflation. 

Although spectral helicity density can not be directly measured, for completeness, 
we give the expression for the spectral helicity density:
\begin{align}\label{eq:helicityPowerSpectrum}
 \frac{d\rho_h}{d\rm{ln}k} &=  \frac{1}{ 2\pi^2} \frac{k^4}{  a^3(\eta(\tau))}
\left(  \,\,\,\,   \left|   C(\tau) \right|^2 \, k^{\frac{1}{2\alpha}} 
  + \left| C_2 \frac{\mathcal{F}(\tau)^{-1} }{\pi} 
  \Gamma \left( \frac{1}{2\alpha}  \right)  \right|^2   \,
  \left( k \,   \tau^4 \right)^{ - \frac{1}{2\alpha} }  \right).
\end{align}

\subsection{Comparison of our model with scalar-field ($f^2(\phi)\, F\tilde{F}$) coupling models}

Often in the literature, the breaking of conformal invariance
of the electromagnetic action is through the non-minimal coupling of the 
electromagnetic field ($f^2(\phi)\, F\tilde{F}$) with scalar field (possibly inflaton). For a suitable choice of the coupling parameters, it has been shown that a sufficient amount of large-scale magnetic fields can be generated~\cite{2001-Vachaspati-PRL,2003-Caprini.etal-PRD,2005-Campanelli-Giannotti-PRD,2018-Sharma.Subramanian.Seshadri.PRD,2009-Caprini.Durrer.Fenu-JCAP,2009-Campanelli-IJMPD,2019-Shtanov-Ukr.PJ}. 

In contrast, our model does not rely on the fine-tuning of the extra coupling parameter to the electromagnetic field and depends on the background quantities through the Riemann tensor. Due to this, the mode functions \eqref{eq:eom_helicity} contain higher derivatives of ${\cal H}$ compared to scalar field coupled models. In Appendix \ref{app:slow-roll-case}, we use a naive approximation of our model for power-law and slow-roll inflation, which effectively ignores higher-derivatives of ${\cal H}$. Under this approximation, we show that the model leads to a blue-tilt spectrum. {Thus, the presence of a higher-derivative of ${\cal H}$ leads to the red-tilt.} This is an important difference between our model compared to scalar-field coupled models. 

To further understand this, we define the overall coupling function in our model by a dimensionless coupling function:
\begin{align}
I = \frac{ R_{\mu\nu}\,^{\sigma\gamma} }{ M^2 } \qquad 
\end{align}
In the flat FRW line-element, Riemann tensor $R_{\mu\nu}\,^{\sigma\gamma} \sim \frac{{a^{\prime}}^2}{a^4}$ and $ \sim \frac{a^{\prime\prime}}{a^3}$. Let us now compare our model with the two specific forms of scalar-field coupled models~ \cite{2011-Durrer.Hollenstein.Jain-JCAP,2018-Sharma.Subramanian.Seshadri.PRD}.

In our model, for power-law inflation, the coupling function $I$ is
\begin{align}\label{eq:coupl-I-constant_M}
 I \sim \frac{ {a^{\prime}}^2 }{M^2 a^4}~~\left(\mbox{or}~\frac{ {a^{\prime\prime}} }{M^2 a^3}\right)
 \propto \frac{1}{\eta^{2\beta + 4}} \propto \eta^{\delta} \qquad (  \delta > 0  )    
\end{align} 
As mentioned above, for the scalar field coupled models, many authors have used different forms of $f^2(\phi)$: 
\begin{equation}
    f_1(\phi) \propto  e^{\frac{\phi}{m_P}}~;~
f_2(\phi) \propto a^2     
\end{equation}
Note that $f_1$ was used in Ref.  \cite{2011-Durrer.Hollenstein.Jain-JCAP} while 
$f_2$ was used in Ref. \cite{2018-Sharma.Subramanian.Seshadri.PRD} . In both the cases, the coupling function is of the form:
\begin{equation}
\label{eq:coupl-f-constant_M}
  f(\phi(\eta)) \propto \frac{1}{\eta^\sigma} \qquad ( \sigma > 0 )      
  \end{equation}
We want to make the following remarks regarding the two coupling forms  \eqref{eq:coupl-I-constant_M} and \eqref{eq:coupl-f-constant_M}:  First, the functional form of the coupling function in our case is different compared to the scalar field coupling models. Since $\eta$ is large at early times, the Riemann coupling term contributes significantly at early times, and hence, the modes that leave the horizon at early times will have large helicity. In the scalar field coupled models, the modes generated at early times will not have significant helicity modes. In contrast, the modes generated close to the end of inflation will have significant helicity. This provides a qualitative understanding of the red-tilt power-spectrum in our case. Second, since most of the helical modes are generated at early times, unlike in Ref.  \cite{2018-Sharma.Subramanian.Seshadri.PRD}, the generated helical fields are not sensitive to the reheating dynamics. Thus, 
the helical modes generated evolve similar to the scalar and tensor perturbations generated during inflation. Today's observable scales (in the CMB and LSS) span roughly three orders in the comoving wavenumber $k$. The largest observable wavelength $\lambda_{\rm max}$, associated with the wavenumber $k_{\rm max}$, corresponds to the horizon radius. For a model with 60 e-foldings of inflation,  the observable cosmological wavelengths exit the Hubble radius around 30 e-foldings before the end of inflation. This will also apply to the helical modes generated in our model. Third, the total energy density of the helical fields in our model is larger compared to the scalar-field coupled models. Specifically, the energy density of the helical fields in our model is at least of an order of magnitude larger than for the coupling function $f_2(\phi)$ in Ref. \cite{2018-Sharma.Subramanian.Seshadri.PRD}. However, as shown in the previous subsection, our model is free from the backreaction problem for a range of scale-factor during inflation.

\subsection{Estimating the strength of the helical magnetic fields}

As mentioned above, the model generates helical fields around $30$ e-folding before the end of inflation. To estimate the current value of the helical fields, we assume instantaneous reheating, and the Universe becomes radiation dominated after inflation.  
Due to flux conservation, the magnetic energy density will decay as ${1}/{a^4}$, i. e.:
\[
\rho_{B}(0) = \rho_B^{(f)} \left(  \frac{a_f}{a_0} \right)^4
\]
where $a_0$ is the present day scale-factor, $\rho_B^{(f)}$ and $a_f$ refer to the magnetic energy density and the scale-factor at the end of inflation, respectively. Using the entropy conservation i.e., $g \, T^3 \, a^3 = constant$ where $g$ refers to the effective relativistic degrees of freedom and $T$ is the temperature of the relativistic fluid, we get $ {a_0}/{a_f} \approx 10^{30} \left( {H_f}/{10^{-5}  M_{\rm{Pl} } }  \right)^{1/2}$~\cite{2016-Subramanian-Arxiv}. 

Using the fact that the relevant modes exited Hubble radius around 30 e-foldings of inflation, with energy density $\rho_B \approx 10^{64} \rm{GeV}^4$, the primordial helical fields at ${\rm GPc}$ scales is:
\begin{align}
B_0 \approx 10^{-20} \rm{G} 
\end{align}
where we have used $1 G = 1.95 \times 10^{-20} \rm{GeV}^2 $ and $H_f=10^{-5} M_{\rm{Pl}}$ is the Hubble parameter during inflation. Our model predicts the following primordial helical fields that re-entered the horizon at two different epochs:
\[
\left. B \right|_{50~{\rm MPc}} \sim 10^{-18}~G~(z \sim 20)~; 
~ \left. B \right|_{1~{\rm MPc}} \sim 10^{-14}~G~(z \sim 1000)\, .
\]
Thus, the model generates sufficient primordial helical magnetic fields at all observable scales.

\section{Conclusions and Discussions}
\label{sec:conc}

We have proposed a viable scenario for the generation of
helical magnetic fields during inflation, which does not require coupling to the scalar field. The generation of the helical fields is
due to the coupling of the electromagnetic fields with the Riemann tensor. To our knowledge, Riemann tensor coupling has not been discussed in the literature to generate helical fields.

The model has many key features: First, it does not require the coupling of the electromagnetic field with the scalar field. Hence, there are no extra degrees of freedom and will not lead to a strong-coupling problem. Second, the conformal invariance is broken due to the coupling to the Riemann tensor. Since the curvature is large in the early Universe, the coupling term will introduce non-trivial corrections to the electromagnetic action. However, at late-times, the new term will not contribute, and the theory is identical to standard electrodynamics\footnote{For instance, at the current epoch, $H_0 \sim 10^{-44} GeV$ and hence, the Riemann coupling term will only contribute in the early Universe and not in the late Universe.}. Third, the power spectrum of the helical fields generated has a slight red-tilt for slow-roll inflation. This is different compared to the scalar field coupled models where the power-spectrum has a blue-tilt. We have also identified the reason for this difference. 
Fourth, our model is free from backreaction for a range of scale-factor during inflation. This is different from other models where a specific form of coupling function is chosen to avoid any back-reaction~\cite{2018-Sharma.Subramanian.Seshadri.PRD}. Interestingly, our model generates the magnetic field of strength $10^{-18} G$ and $10^{-14} G$ over scales $\sim 50 \rm{Mpc}$ and $\sim 1 \rm{Mpc}$, respectively.

In this work, we did not discuss the generation of non-helical magnetic fields. The generation of the non-helical magnetic field with Riemann coupling has been discussed in the seminal paper by Turner and Widrow~\cite{1988-Turner.Widrow-PRD}. An analysis including parity preserving term can be done straightforwardly, and we can obtain total energy density of the non-helical ($\rho_{B}$) and helical energy density ($\rho_H$). As shown recently, the two energy densities must satisfy the realizability condition~\cite{2014-Kahniashvili.etal-PRD}, i. e.,  $\rho_H \leq 2 \, \xi_M \rho_{B}$, where $\xi_M$ is the magnetic correlation length. Assuming that the non-helical and helical power spectra are a power-law:
\[
 P_{B} = A_{B} k^{n_{B}}, \quad P_{H} = A_{H} k^{n_{H}}   \, ,    
\]
{for maximal helicity, it was shown that the helical magnetic fields must have red-tilt.} More specifically, for the WMAP nine-year data, using the cross power-spectrum between the temperature and B-mode polarization they set $95\%$ confidence level upper limit on the helicity amplitude to be $10 \rm{nG}^2 \, \rm{Gpc} $ for the helical spectral index $n_H = -1.9$ and for a cosmological magnetic field with effective field strength of $3~\rm{nG}$ and $n_B = -2.9$. PLANCK 2015 data placed constraints on the strength for causally generated magnetic fields with spectral index $n_B = 2$ and fields with almost scale-invariant spectrum with $n_B = -2.9$ are $B_{1\rm{Mpc}} < .011~\rm{nG}$ and  $B_{1\rm{Mpc}} < 0.9~\rm{nG}$ at $95\%$ confidence level~\cite{2015-Planck-PMF}. Thus, the PLANCK 2015 data also prefers $n_H$ to be negative. With improved B-mode polarization measurements, helicity modes can be better constrained and put our model's prediction to test with the CMB data. We hope to address this soon. 

The perturbations equations \eqref{eq:equation_of_motion} contain second-order derivatives of $\mathcal{H}$. Since, ${\cal H}$, and ${\cal H}''$ are different for inflation and bounce models~\cite{2020-Nandi-PLB}, the helicity modes may provide signatures to distinguish the two paradigms. This is currently under investigation.

\begin{acknowledgments}
The authors thank Debottam Nandi, Archana Sangwan, T. R. Seshadri, Ramkishor Sharma, and K. Subramanian for useful discussions. The authors thank the anonymous referee for raising some points which clarified important issues in the work. The MHRD fellowship at IIT Bombay financially supports AK. This work is supported by the ISRO-Respond grant. 
\end{acknowledgments}

\appendix
\section{Series expansion of the Bessel function}\label{app:Series-exp}

For completeness, in this appendix, we obtain the series expansion of the Bessel functions to the leading order terms~\cite{2010-Olver.etal-Book}. It is useful to define the following quantity:
\begin{align}\label{eq:floor-functions}
F(\tau)  = \exp\left[\frac{i \pi}{\alpha} \left\lfloor \frac{\pi -\arg (\tau)-\arg
   \left(\frac{ \sqrt{k} \, \varsigma}{\alpha }\right) }{2 \pi }\right\rfloor \right]
\end{align}
where $\left\lfloor \cdots \right\rfloor$ represents the floor function. Up to leading order, the series expansion for the Bessel functions are:
\begin{align}
& J_{  \frac{1}{2 \alpha}} \left( \frac{\varsigma \, \sqrt{k} }{\alpha} \tau \, \right) =  \frac{F(\tau) }{\Gamma \left( 1+ \frac{1}{2\alpha}  \right) } \left( \frac{\varsigma \, \sqrt{k} }{2\alpha} \right)^{\frac{1}{2\alpha}}
   \tau^{ \frac{1}{2\alpha} }\left( 1 - \frac{\varsigma^2 k \tau^2}{2 \alpha (1+2\alpha) }+O\left(\tau^3\right)\right), \\
& Y_{  \frac{1}{2 \alpha}} \left( \frac{\varsigma \, \sqrt{k}  }{\alpha} \tau \, \right) = \frac{F(\tau) }{\pi} \left( \frac{\varsigma \, \sqrt{k} }{2\alpha} \right)^{\frac{1}{2\alpha}}
  \Gamma \left( - \frac{1}{2\alpha}  \right) \rm{cos}\left( \frac{\pi}{2\alpha} \right) \left(  - \tau^{\frac{1}{2\alpha}} +  \frac{ k \, \varsigma^2 \, \tau^{2 + \frac{1}{2\alpha}}}{2 \alpha (1 + 2\alpha) }+O\left(\tau^3\right)\right)
  \nonumber \\
  &~~~~~~~~~~~~~~~~~~~~~+ \frac{F(\tau)^{-1} }{\pi} \left( \frac{2\alpha}{ \varsigma \, \sqrt{k} } \right)^{\frac{1}{2\alpha}}
  \Gamma \left( \frac{1}{2\alpha}  \right)  \left( -  \tau^{ - \frac{1}{2\alpha} } + \frac{ k \, \varsigma^2 \,\tau^{2 - \frac{1}{2\alpha}}}{2 \alpha (-1+2\alpha) }+O\left(\tau^3\right)\right), \\
 %%% 
& J_{  \frac{1}{2 \alpha}} \left( -\frac{i \varsigma \, \sqrt{k} }{\alpha} \tau \, \right) 
=  \frac{\tilde{F}(\tau) }{\Gamma \left( 1+ \frac{1}{2\alpha}  \right) } \left( - \frac{ i \,\varsigma \, \sqrt{k} }{2\alpha} \right)^{\frac{1}{2\alpha}}
  \tau^{ \frac{1}{2\alpha} } \left(  1 + \frac{\varsigma^2 k \tau^2}{2 \alpha (1+2\alpha) }+O\left(\tau^3\right)\right) \\
& Y_{  \frac{1}{2 \alpha}} \left(-i \frac{\varsigma \, \sqrt{k}  }{\alpha} \tau \, \right) = -\frac{\tilde{F}(\tau) }{\pi} \left( -\frac{ i \, \varsigma \, \sqrt{k} }{2\alpha} \right)^{\frac{1}{2\alpha}}
  \Gamma \left( - \frac{1}{2\alpha}  \right) \rm{cos}\left( \frac{\pi}{2\alpha} \right) \tau^{\frac{1}{2\alpha}} \left(   1 +  \frac{ k \varsigma^2\tau^2 }{2 \alpha (1 + 2\alpha) }+O\left(\tau^3\right)\right) \nonumber
\end{align}
\begin{align}  
&~~~~~~~~~~~~~~{} - \frac{\tilde{F}(\tau)^{-1} }{\pi} \left( -\frac{2\alpha}{ \, i \, \varsigma \, \sqrt{k} } \right)^{\frac{1}{2\alpha}}
  \Gamma \left( \frac{1}{2\alpha}  \right)  \tau^{ - \frac{1}{2\alpha} }  \left( 1 + \frac{ k \, \varsigma^2 \,\tau^2}{2 \alpha (-1+2\alpha) }+O\left(\tau^3\right)\right).
\end{align}
At leading order, the helicity modes are:
\begin{align}\label{Appeq:A+}
A_+(\tau,k) &= F(\tau) \, \left( \frac{\varsigma \, \sqrt{k} }{2\alpha} \right)^{\frac{1}{2\alpha}} \left[  \frac{C_1 }{\Gamma \left( 1+ \frac{1}{2\alpha}  \right) } 
   -  \frac{C_2 }{\pi}  \Gamma \left( - \frac{1}{2\alpha}  \right) \rm{cos}\left( \frac{\pi}{2\alpha} \right) \right] \nonumber \\
 &-~~ C_2 \frac{F(\tau)^{-1} }{\pi} \left( \frac{2\alpha}{\varsigma \, \sqrt{k} } \right)^{\frac{1}{2\alpha}}
  \Gamma \left( \frac{1}{2\alpha}  \right)  \tau^{ - \frac{1}{\alpha} }   \\
% \end{align} 
%  
%\begin{align}
\label{Appeq:A-}
A_-(\tau,k) &=\tilde{F}(\tau) \left( -i \frac{\varsigma \, \sqrt{k} }{2\alpha} \right)^{\frac{1}{2\alpha}}  \left[ \,  \frac{C_3 }{\Gamma \left( 1+ \frac{1}{2\alpha}  \right) }  -  \frac{C_4}{\pi}
\Gamma \left( - \frac{1}{2\alpha}  \right) \rm{cos}\left( \frac{\pi}{2\alpha} \right)  \right] \nonumber\\
&{}-~~C_4 \frac{\tilde{F}(\tau)^{-1} }{\pi} \left( -\frac{2\alpha}{ i \varsigma \, \sqrt{k} } \right)^{\frac{1}{2\alpha}}
  \Gamma \left( \frac{1}{2\alpha}  \right)  \tau^{ -\frac{1}{\alpha} } 
\end{align}
\begin{figure}[!hbt]
	\centering
	\includegraphics[width=0.65\textwidth]{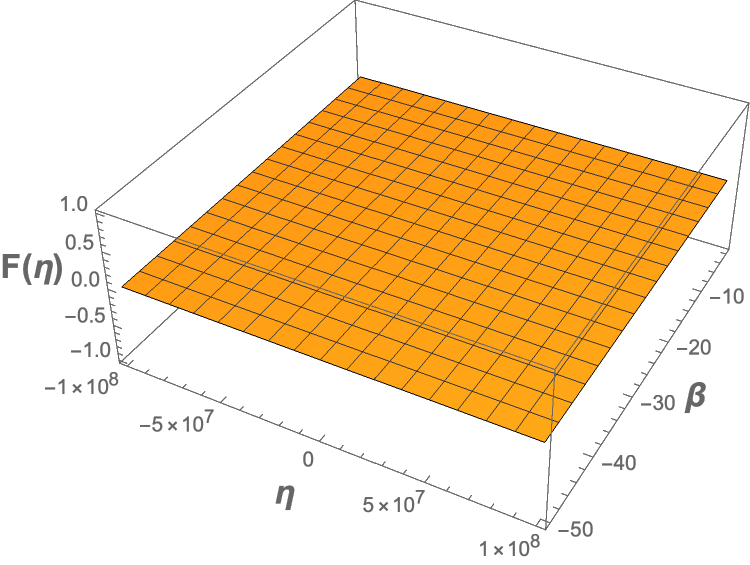}
	\caption{Plot of the Floor function for the range of $-10^8 < \eta < 10^8$ and $-50 < \beta < -2$.}
	\label{fig:Plot}
\end{figure}

It is convenient to define the following quantities:
\begin{align}\label{eq:quantities-F}
\mathcal{F}(\tau) &= F(\tau) \, \left( \frac{\varsigma }{2\alpha} \right)^{\frac{1}{2\alpha}}, \\
\label{eq:quantities-Ftilde}
\tilde{\mathcal{F}}(\tau) &= \tilde{F}(\tau) \, \left( -i\frac{\varsigma }{2\alpha} \right)^{\frac{1}{2\alpha}}, \\
\label{eq:quantities-C}
C(\tau) &= F(\tau) \, \left( \frac{\varsigma }{2\alpha} \right)^{\frac{1}{2\alpha}} \left[  \frac{C_1 }{\Gamma \left( 1+ \frac{1}{2\alpha}  \right) } 
   -  \frac{C_2 }{\pi}  \Gamma \left( - \frac{1}{2\alpha}  \right) \rm{cos}\left( \frac{\pi}{2\alpha} \right) \right], 
 \end{align} 
\begin{align}  
   \label{eq:quantities-Ctilde}
\tilde{C}(\tau) & = \tilde{F}(\tau) \left( -i \frac{\varsigma}{2\alpha} \right)^{\frac{1}{2\alpha}}  \left[ \,  \frac{C_3 }{\Gamma \left( 1+ \frac{1}{2\alpha}  \right) } 
-\frac{C_4}{\pi}
\Gamma \left( - \frac{1}{2\alpha}  \right) \rm{cos}\left( \frac{\pi}{2\alpha} \right)  \right]
\end{align}
\ref{fig:Plot} is the 3-D plot of the floor function for the range of values of $\eta$ and $\beta$ that are consistent with generic models of inflation. As can be seen from the plot, the floor function is zero in the range of interest. Due to this reason, we have suppressed the time dependence in above quantities i.e., $\mathcal{F}(\tau) = \mathcal{F},\tilde{\mathcal{F}}(\tau) = \tilde{\mathcal{F}}$ and $C(\tau) = C, \tilde{C}(\tau) = \tilde{C}$. We have used this in computing the energy densities in Sec. \eqref{sec:Helical}. 

%
%%%%%%%%  S E C T I O N %%%%%%%%%%
%
%
%%%%%%%%%%%%%   Case M = H %%%%%%%%%%%%%%%%%
\section{Power spectrum for slow-roll inflation}
\label{app:slow-roll-case}

In Sec. \eqref{sec:Helical}, we obtained the power-spectrum for the helical fields in the power-law inflation. In this section, we obtain the power-spectrum for slow-roll inflation. 

To do that, we first obtain the power-spectrum for the power-law inflation by 
assuming that $M$ is slowly varying and is related to the Hubble parameter 
$H$, i. e., $M \sim H \sim  \mathcal{H}/{a}$.

\subsection{Power law inflation}
The equation of motion (\ref{eq:eom_helicity}) for power inflation (\ref{eq:powerLaw}) is: 
\begin{align}\label{eq:eom_helicity-PL-beta}
A_h^{\prime\prime} + \left[  k^2 + \frac{8hk\xi}{\eta} \right] \, A_h= 0.
\end{align}
where $\xi = \frac{\beta (\beta + 2)}{(\beta + 1) }$, solution of the above equation for super horizon modes are given by
\begin{align}\label{eq:+sup-mod-M=H}
A_+(\eta, k) &= 2 \sqrt{2 k \xi \eta} \left[ \, D_1 J_1\left(4 \sqrt{ 2 k \xi \eta} \right)+2 i D_2 Y_1\left(4 \sqrt{2 k \xi \eta} \, \right)\right]\\
\label{eq:-sup-mod-M=H}
A_-(\eta, k) &= -2 \sqrt{2 k \xi \eta} \left[  \, D_3 I_1\left(4 \sqrt{2 k \xi \eta} \right)+ 2 D_4 \,  K_1\left(4 \sqrt{2 k \xi \eta} \right)\right] \, ,
\end{align}
{   where $D_1,D_2(D_3,D_4)$ are the arbitrary constants, we assume these are of the same order of $C_1,C_2(C_3,C_4)$ respectively.} As was shown earlier, we can set $\left| A_-(\eta, k)   \right| = 0$, and positive mode can be approximated by power series at leading order as 
%\begin{align}\label{eq:+sup-mod-M=H-series}
%A_+(\eta, k) &= 8D_1 \xi k \eta - D_2\,  \left[\,\, 64 F(\eta,k) \xi k \eta - \frac{2i}{\pi} \,\, \right].
%\end{align}
%
\begin{align}\label{eq:+sup-mod-M=H-series}
A_+(\eta, k) &= 8D_1 \xi k \eta + D_2\, \frac{2i}{\pi} \, .
\end{align}
where we have used $F(\eta,k) = 0$ in the above eq. (\ref{eq:+sup-mod-M=H-series}), therefore spectral magnetic energy density at horizon exit is given by
%
%
%\begin{align}\label{eq:spectral_rhoB-case-M=H}
%\left.  \frac{d\rho_B}{d\rm{ln}k} \right|_{k_* \eta_* \sim 1} &=  \frac{( - \eta_0)^{4\beta + 4}}{\left( 2\pi \right)^2} k_*^{4\beta + 9}  \left( \,\, \left| 8D_1 \xi - D_2 \left( 64 F(\eta_*,k_*) \xi - \frac{2i}{\pi} \right)  \right|^2 \right.
%  \nonumber\\
%  &{}
% \left.+ \left| 4D_3 \xi  + D_4 \left( 32 \pi i F(\eta_*,k_*) \xi + 1   \right) \, \, \right|^2  \,\,\,\, \right).
%\end{align}
%
%
\begin{align}\label{eq:PowSpec_B_M=H0}
\left.  \frac{d\rho_B}{d\rm{ln}k} \right|_{k_* \eta_* \sim 1} =  \frac{( - \eta_0)^{4\beta + 4}}{\left( 2\pi \right)^2} k_*^{4\beta + 9}  \left( \,\, \left| 8D_1 \xi + D_2 \frac{2i}{\pi}  \right|^2  \,\,\, \right)
\end{align}
Unlike the exact calculation in Sec. \eqref{sec:Helical}, this approximation leads to scale invariant spectrum for $\beta = - \frac{9}{4}$. Substituting the values of the coefficients $D_1$ and $D_2$ from eq.(\ref{eq:Coeff-value}), we obtain:
 \begin{align}\label{eq:PowSpec_B_M=H}
\left.  \frac{d\rho_B}{d\rm{ln}k} \right|_{k_* \eta_* \sim 1} \approx 10^{59} \rm{GeV}^4= 10^3 \mathcal{H}^4
\end{align}
Thus, $R \sim 10^{-9}$. This implies that the above approximation leads to a reduction in the helical field power-spectrum. We will now use this procedure to evaluate the power-spectrum in slow-roll inflation. 
 %
%
%%%%%%%%%%  S E C T I O N %%%%%%%%%%%%%%%
%
%
\subsection{ Slow roll inflation}
The slow-roll parameter $\epsilon$ in terms of $\mathcal{H}$ is defined by 
\begin{align}
\epsilon = 1 - \frac{\mathcal{H}^{\prime}}{ \mathcal{H}^2 }
\end{align}
A necessary condition for inflation is  $\epsilon< 1$. For the leading order slow-roll, we have:
\begin{equation}
\mathcal{H} \approx - \frac{ (1+\epsilon) }{\eta};~\mathcal{H}^{\prime\prime} \approx -\frac{2(1+\epsilon)}{\eta^3};
\end{equation}
Substituting these in Eq.~(\ref{eq:eom_helicity}), we have:
\begin{align}\label{eq:eom_helicity_sr-epsilon}
A_h^{\prime\prime} +\left[  k^2 - \frac{16kh}{\eta} \frac{\epsilon}{1+\epsilon}\,  \right] \, A_h \approx 0.
\end{align}
Note that the above equation (\ref{eq:eom_helicity_sr-epsilon}) can also be obtained by substituting $\beta = -2-\epsilon$ in Eq.~(\ref{eq:eom_helicity-PL-beta}) where the expression of $\xi$ will be $\xi = -\frac{2\epsilon}{1+\epsilon}$. Hence, the super-horizon mode solution can be obtained by substituting $\xi = -\frac{2\epsilon}{1+\epsilon}$ in eq.(\ref{eq:+sup-mod-M=H}) and (\ref{eq:-sup-mod-M=H}). Thus after setting $ \left| A_-(\eta,k) \right|= 0$, spectral magnetic energy density for slow roll case will have the form
\begin{align}\label{eq:PowSpec_B-M=H-slowroll}
  \left.  \frac{d\rho_B}{d\rm{ln}k} \right|_{k_* \eta_* \sim 1}  &= {  \frac{1}{\left( 2\pi \right)^2} \frac{k_*^{1 - 4\epsilon}}{  (-\eta_0)^{4\epsilon + 4} } \left( \,\, \left| -\frac{16\epsilon }{1+\epsilon}D_1  +  \frac{2i}{\pi} D_2  \right|^2  \,\,\, \right)  }
\end{align}
Unlike the exact calculation in Sec. \eqref{sec:Helical}, this approximation leads to blue-tilt spectrum. Thus, this is not a  approximation to obtain power-spectrum for helical magnetic fields. 


%\bibliography{references_all}
\end{document}